\documentclass[letterpaper,10pt]{article} 

\usepackage{opticameet3} 
\usepackage{titlesec}
\titlespacing*{\section}
{0pt}{1ex plus .4ex minus .2ex}{1ex plus .2ex}
\providecommand{\myparab}[1]{\vspace{1pt}\noindent\textbf{#1} }

\newcommand\authormark[1]{\textsuperscript{#1}}
\usepackage[font=small,labelfont=bf]{caption}
\usepackage{amsmath,amssymb}
\usepackage{graphicx}
\usepackage{wrapfig}
\usepackage{subcaption}
\usepackage[colorlinks=true,bookmarks=false,citecolor=blue,urlcolor=blue]{hyperref} 
\newcommand{\name}{PipSwitch}

\begin{document}

\title{\name: A Circuit Switch Using Programmable Integrated Photonics}
\vspace{-2em}


\author{Eric Ding,\authormark{1,*} and Rachee Singh,\authormark{1}}

\address{\authormark{1} Cornell University}
\email{\authormark{*}ericding@cs.cornell.edu}
\vspace{-1em}

\providecommand{\vs}{vs. }
\providecommand{\ie}{\emph{i.e.,} }
\providecommand{\eg}{\emph{e.g.,} }
\providecommand{\aka}{\emph{aka} }
\providecommand{\cf}{\emph{cf.,} }
\providecommand{\resp}{\emph{resp.,} }
\providecommand{\etal}{\emph{et al. }}   
\providecommand{\etc}{\emph{etc.}}      
\vspace{-1em}
\begin{abstract}
We present an optical circuit switch design for programmable integrated photonics (PIPs). Our solution finds the correct and optimal set of matchings that provides all-to-all network connectivity and demonstrates scalability to 32 ports.
\end{abstract}

\section{Introduction}

Programmable integrated photonics (PIP) is an emerging field that aims to create reconfigurable and versatile optical circuits on a chip using a mesh of waveguides, tunable beam couplers, and optical phase shifters. The optical hardware can be configured to perform a wide variety of functionalities, such as signal processing, neuromorphic computing, and quantum information processing using software \cite{bogaerts2020programmable}.

We focus on a novel application of PIP: developing an optical circuit switch. PIP-based switches could offer advantages similar to existing optical switches (\eg MEMs switch) over traditional electrical switches, including energy efficiency and low latency since there is no queuing and header processing in optical switches. Additionally, PIP devices can provide several benefits over fixed-function optical devices, including greater flexibility through software-defined operations, enabling the on-demand implementation of various network topologies, and compatibility with existing CMOS manufacturing techniques \cite{bogaerts2020programmable}. However, PIP's complex mesh topology poses challenges in making optimal switching decisions and maximizing switch connectivity \cite{kerchove2023automated}.

We present \name, a PIP-based circuit switch that (1) provides all-to-all network connectivity by finding the correct and optimal set of matchings within PIP constraints, (2) scales adequately to hundreds of PIP cells, supporting up to 32 ports with $9\times9$ PIP mesh size, and (3) achieves an end-to-end electrical-optical implementation.

\begin{figure}[h!]
\vspace{-1em}
  \centering
    \includegraphics[width=0.82\textwidth]{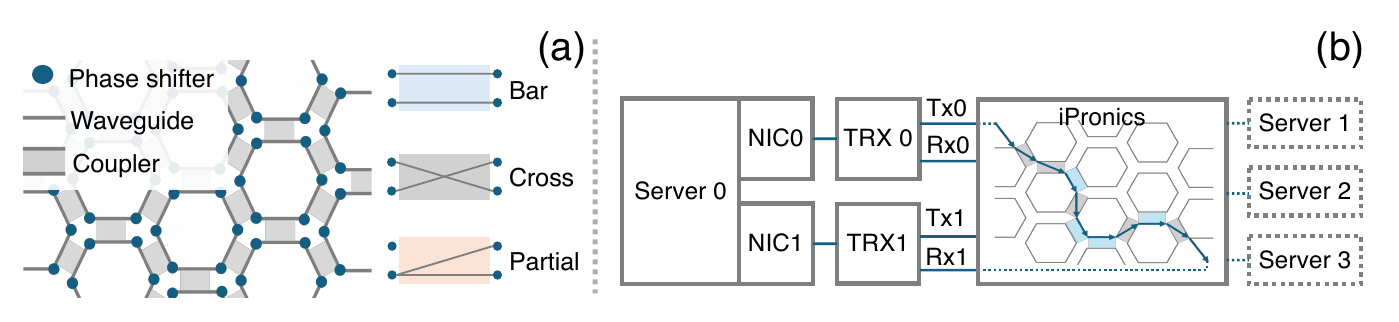}
  \caption{\small{(a): Programmable integrated photonics. (b): Two network interfaces were connected via two routes configured in iPronics (only one is shown). More servers can be connected to construct a rack.}}
  \label{fig:pip1}
\end{figure}

\section{Designing Circuit Switches Using PIPs}
\label{sec:s1}
\begin{wraptable}{r}{0.45\textwidth}
    \vspace{-1mm}
    \footnotesize
    \centering
    \caption{\small{End-to-end measurements of latency and bitrate through the PIP mesh using the \emph{Ping} and \emph{iPerf} tool. Interface 0 initiates measurements. Route length is the number of PUCs on the Tx0-Rx1 route.}}
    \begin{tabular}{p{0.5cm} p{2.8cm} p{1cm} p{1cm}}
        \hline
        Route length & \emph{Ping} min/avg/max/mdev RTT (ms) & \emph{Ping} packet loss (\%) & \emph{iPerf} bitrate (Gb/s)\\
        \hline
        9	& 0.176/0.261/0.287/0.019	& 0	    & 9.41 \\
        13	& 0.185/0.248/0.402/0.018	& 0	    & 9.41 \\
        17	& 0.169/0.243/0.299/0.023	& 0	    & 9.41 \\
        19	& 0.152/0.231/0.268/0.017	& 1.14   & 3.08 \\
        21	& 0.178/0.242/0.314/0.013	& 2.84	    & 2.38 \\
        23	& 0.169/0.257/0.820/0.055	& 22.73	& 0 \\
        25	& 0.177/0.247/0.279/0.019	& 81.25	& 0 \\
        27	& n/a                       & 100	& 0 \\
        \hline
   \end{tabular}
   \vspace{-2mm}
   \label{tb:measurement}
\end{wraptable}
A PIP device typically consists of a 2D mesh of programmable unit cells (PUCs) (Fig. \ref{fig:pip1}(a)). The popular hexagonal mesh design enables recirculation of light to synthesize different photonic circuits \cite{lopez2019programmable}. Every hexagon is called a cell, which is formed by six surrounding PUCs. A PUC is a $2\times2$ optical gate based on a tunable coupler and thermo-optical phase shifters. Each individual PUC can be programmed into one of the three states: bar, cross, and partial state to realize any $2\times2$ complex unitary transfer matrix. 
In the following discussion, we only consider the bar and cross states and leave multi-casting behavior achieved by the partial state for future work.

We used a commercial PIP offering from iPronics\cite{lopez2019programmable} to study the end-to-end data transmission characteristics on a PUC mesh. Using these experiments, we determine (1) sustainable data rates of optical routes in the mesh and (2) optical route lengths that achieve lossless data transmission. For this, we connected two network ports of a Dell server running Ubuntu 22.04 to iPronics using 10G optical transceivers (Fig.~\ref{fig:pip1}(b)). We programmed the mesh to establish two routes between these ports, each route connecting the Tx of one port to the Rx of the other. The two routes enabled full-duplex communication between two Linux network namespaces. We changed the length of the circuit between Tx0 and Rx1, and studied the impact of the route length on network performance (Table \ref{tb:measurement}). Beyond the route length of 17 PUCs, we observed packet drops due to diminished signal power. We also measured the hardware configuration time. Configuring one PUC takes 47.189 ms on average. And there is a linear relationship ($k = 0.005$) between total hardware configuration latency and route length measured by the number of PUCs.

We summarize properties and constraints of the PIP hardware that influence the design of {\name}:
\textbf{P1.} \label{size} There is a positive correlation between the hardware configuration time and the total length of the routes. 
\textbf{P2.} \label{weight}Every PUC coupler arm (waveguide) incurs propagation, bend, and insertion signal loss.
\textbf{C1.} \label{len} The length of the route has an upper bound (follows from Property \textbf{P2}).
\textbf{C2.} \label{conflict} There should be no PUC state conflict. A PUC can be shared by two different routes that use different coupler arms, but it can only be in one state (either bar or cross state). Also, one route cannot use both the bar arm and cross arm of a coupler.
\textbf{C3.} \label{noref} Any coupler arm cannot be shared. Two routes cannot use the same arm, and one route cannot traverse the same arm twice. 

\section{\name: Switching Problem Formulation}

To implement switching functionality, {\name} needs to identify a matching between ports of the mesh to implement the desired network topology. We formulate this as an Integer Linear Program (ILP) with the goal of finding an optimized matching among a given set of ports while respecting PIP hardware properties and constraints. We define a variable $x_{r, e} \in \{0, 1\}$ for every combination of network route (source-drain pair) $r \in R$ and edge (PUC coupler arm) $e \in E$, where $R$ is the set of all routes and $E$ is the set of all edges. $x_{r, e} = 1$ means route $r$ traverses $e$. Note that we could uniquely determine the state of a PUC based on $x_{r, e}$ that is set to 1. 

\begin{wrapfigure}{r}{0.44\textwidth}
    \vspace{-2em}
  \begin{center}
      \includegraphics[width=0.44\textwidth]{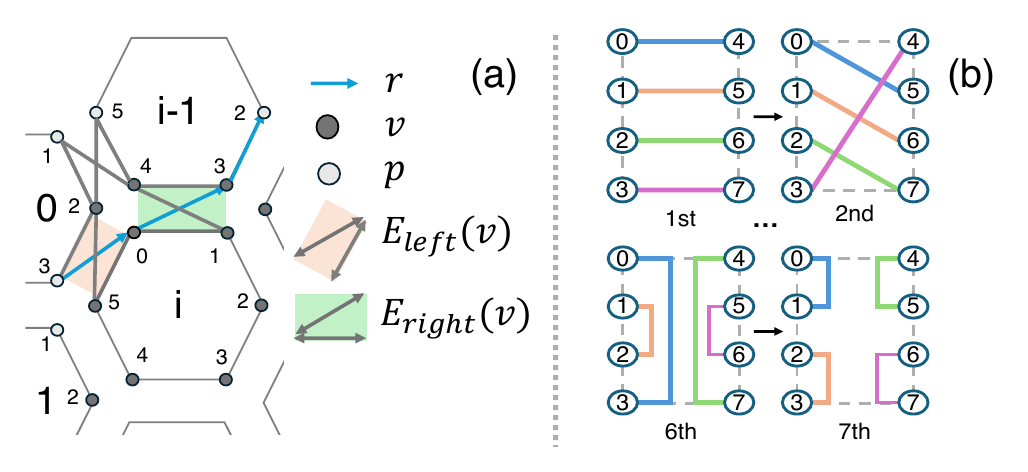}
  \end{center}
  \vspace{-1em}
  \caption{\small{(a): Illustration for the ILP. Every cell has an ID numbered from top to bottom, left to right. Every endpoint has a local ID within the cell. There is one route with (cell 0, port 3) as source and (cell i-1, port 2) as drain. Its length is 3. Integer variables for the blue lines are set to 1. (b): Sample matchings for a rotor switch.}} 
  \vspace{-1em}
  \label{fig:pip2} 
\end{wrapfigure}

The set of endpoints of any $e \in E$ is the union of $V$ and $P$, where $V$ is the set of vertices (phase shifter) within the device and $P$ is the set of ports on the edge of the device (Fig. \ref{fig:pip2}(a)). $R$ is the \textbf{input} to the model in the form of source-drain pairs. $r_s$ and $r_d$ are the source and drain for each $r \in R$, and $r_s, r_d \in P$. The \textbf{output} of the model is the state of each PUC (derived from $x_{r, e}$) which is then programmed into the PIP hardware if there is a solution, or \texttt{nil}. $E(v)$ is the set of edges of which $v$ is the endpoint. $E_{in}(v)$ is the set of incoming edges of $v$, and $E_{out}(v)$ is the set of outgoing edges of $v$. $E_{left}(v)$ is the set of edges on the left side of $v$; $E_{right}(v)$ is the set of edges on the right side (Fig. \ref{fig:pip2}(a)). 
The objective is to minimize the total length of circuits (total number of PUCs) in view of \textbf{P1}:
\setlength{\abovedisplayskip}{0pt} 
\setlength{\belowdisplayskip}{0pt} 
$$
    \textrm{Minimize} \ \sum_{r\in R}\sum_{e\in E} c_e x_{r, e}, \textrm{where $c_e$ is the configuration overhead of $x_{r,e}$.}
$$

The solution is subject to the following constraints:
\setlength{\abovedisplayskip}{0pt} 
\setlength{\belowdisplayskip}{0pt} 
\begin{equation}
    \sum_{e \in E} w_e x_{r, e} \leq L, \ \forall r \in R, \ \textrm{where $w_e$ models the signal loss of each edge (coupler arm)}.
    \label{eq:maxlen}
\end{equation}
\setlength{\abovedisplayskip}{0pt} 
\setlength{\belowdisplayskip}{0pt} 
\begin{equation}
    \sum_{e\in E_{in}(v)} x_{r, e} = \sum_{e\in E_{out}(v)} x_{r, e},\  \forall v \in V,\ \forall r \in R
    \label{eq:kirchoff}
\end{equation}
\begin{equation}
    \sum_{r \in R}\sum_{e\in E_{left}(v)} x_{r, e} \leq 1,\ \sum_{r \in R}\sum_{e \in E_{right}(v)} x_{r, e} \leq 1,\ \forall v \in V
    \label{eq:maxside}
\end{equation}

\begin{equation}
     \sum_{e \in E_{out}(r_s)} x_{r, e} = \sum_{r' \in R}\sum_{e \in E_{out}(r_s)} x_{r', e} = 1, \ \sum_{r' \in R}\sum_{e \in E_{in}(r_s)} x_{r', e} = 0,\ \forall r \in R, \ \textrm{where $r_s$ is the source port of $r$, $r_s \in P$.}
    \label{eq:source}
\end{equation}


\begin{equation}
     \sum_{e \in E_{in}(r_d)} x_{r, e} = \sum_{r' \in R}\sum_{e \in E_{in}(r_d)} x_{r', e} = 1, \ \sum_{r' \in R}\sum_{e \in E_{out}(r_d)} x_{r', e} = 0,\ \forall r \in R, \ \textrm{where $r_d$ is the drain port of $r$, $r_d \in P$.}
    \label{eq:drain}
\end{equation}

\begin{equation}
    \sum_{r \in R}\sum_{e \in E_{in}(p)} x_{r, e} = \sum_{r \in R}\sum_{e \in E_{out}(p)} x_{r, e} = 0,\ \forall p \in P \setminus \{r_d, r_s \ | \ r \in R\}
    \label{eq:port}
\end{equation}

Eq.~\ref{eq:kirchoff} is Kirchhoff's law, stating that the number of incoming flows of a vertex $v \in V$ should be equal to the number of outgoing flows for each route. The number of outgoing flows is equal to 1, whereas the number of incoming flows is 0 for a source port $p \in P$ (Eq.~\ref{eq:source}), and vice versa when $p$ is a drain port (Eq.~\ref{eq:drain}). If $p$ is not associated with a route, the total flow number should be zero (Eq.~\ref{eq:port}). Eq.~\ref{eq:maxside} is derived from \textbf{C2} and \textbf{C3}. For example, only one edge can be selected from $E_{left}$(cell i, vertex 0) = \{(cell 0, port 3)$\xrightarrow{}$(cell i, vertex 0), (cell i, vertex 0)$\xrightarrow{}$(cell 0, port 3), (cell i, vertex 5)$\xrightarrow{}$(cell i, vertex 0), (cell i, vertex 0)$\xrightarrow{}$(cell i, vertex 5)\} in Fig.~\ref{fig:pip2}(a). 

Past work has proposed routing algorithms on a PIP mesh~\cite{jia2018six, lopez2020auto, kerchove2023automated}. However, these algorithms are based on heuristics, they prioritize minimizing computation time over finding optimal solutions. In other words, these algorithms may fail to find a switching solution, \ie a matching among a set of device ports. This limits the feasibility of using PIP devices as circuit switches. \cite{kerchove2023automated} proposed similar ILP models as above. However, our formulation ensures the correctness of the routing solutions due to the following differences compared to \cite{kerchove2023automated}: 

(1) Our experiments (\S\ref{sec:s1}) show that routes with PUCs greater than 17 are lossy. So, we introduce Eq.~\ref{eq:maxlen} to limit the length of individual routes in addition to the objective of minimizing the number of PUCs used in the circuit.
(2) PUC state conflict incurred by one route using both the cross and bar arms of a PUC is ruled out by Eq.~\ref{eq:maxside}. For example, the route (cell i, vertex 5)$\xrightarrow{}$(cell i, vertex 0)$\xrightarrow{}$(cell 0, port 3) in Fig.~\ref{fig:pip2}(a) is infeasible in our solution.
(3) PUC state conflict incurred by multiple routes is ruled out by Eq~\ref{eq:maxside}. For example, route 0 using (cell i, vertex 5)$\xrightarrow{}$(cell i, vertex 0) and route 1 using (cell 0, vertex 3)$\xrightarrow{}$(cell i, vertex 0) is infeasible in our solution.
(4) Eq.~\ref{eq:port} forbids a route to cross disconnected ports (leaving the device and entering the device during propagation).

In addition to correctly solving one matching, we implement a rotor switch using {\name} to achieve all-to-all connectivity among the switch ports.
A naive approach is to cycle through all $O(radix!)$ possible matchings, but the cycle time will quickly explode as the radix number increases. Instead, we use RotorNet's approach, rotating over $O(radix)$ matchings in a cycle so that any pair of ports is guaranteed to get connected within a fixed period of time \cite{mellette2017rotornet}. The degree of an end host in our model, however, is 1, contrary to the rotor switch setup in \cite{mellette2017rotornet} which has a degree of 2, as the route in {\name} is more fine-grained (Tx-Rx pair). Fig.~\ref{fig:pip2}(b) shows 4 of the 7 matchings in a rotor cycle. 

\section{Evaluating Scalability of \name}

We evaluate the radix of \name, \ie how many ports can the switch interconnect, configuration latency, \ie how fast can we reconfigure {\name}'s port matching, and scalability, \ie how does {\name}'s performance change with increase in mesh size?
We assume that the propagation loss is similar across all PUCs. We observe the variance of the hardware configuration latency across all PUCs is small ($std = 3.96 \ ms$).  Thus, we set $c_e$, $x_e$ as 1 for all $e \in E$. The maximum route length constraint is set as $2\times(H + W)$, where $H$ and $W$ are the height and width of the device. We use Gurobi to solve the optimization using the barrier algorithm.

\begin{wrapfigure}{r}{0.44\textwidth}
\vspace{-3em}
  \begin{center}
      \includegraphics[width=0.45\textwidth]{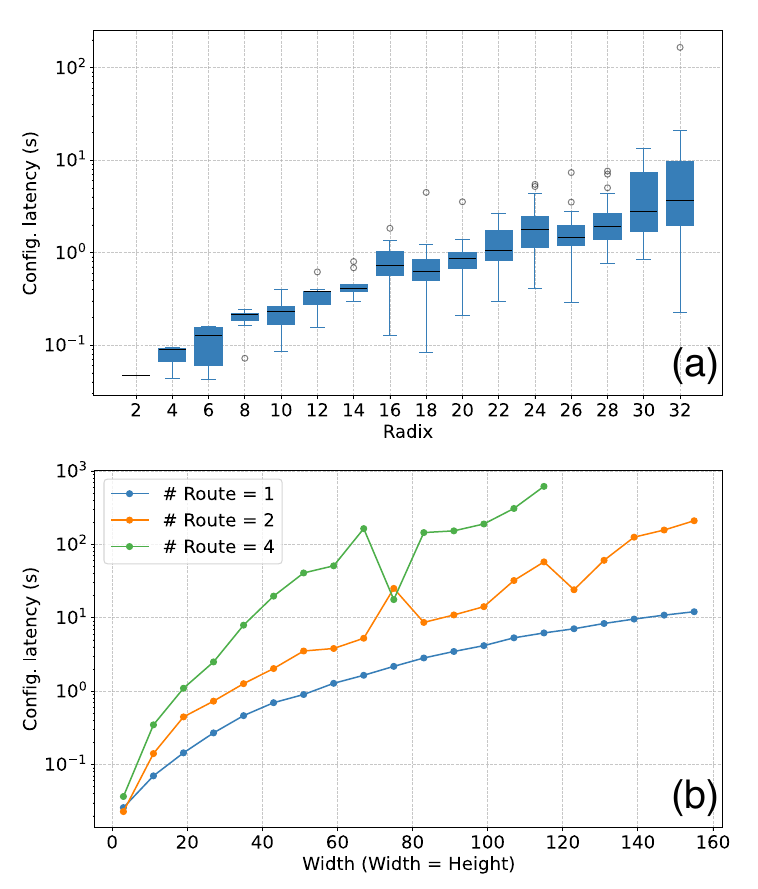}
  \end{center}
    \vspace{-1em}
  \caption{(a): Scaling radix. (b): Configuration latency across different PIP sizes with 1, 2, and 4 routes.}
  \vspace{-1em}
  \label{fig:pip3} 
\end{wrapfigure}

\myparab{Scaling Radix:}
We measure the configuration latency of all rotor matchings given the PIP mesh size ($H = W = 9$) and port positions while varying the radix of the mesh. Fig.~\ref{fig:pip3}(a) shows that {\name} can support up to 32 ports with a $9\times9$ mesh. Exceeding this radix leads to infeasible matchings.

\myparab{Scaling mesh size:}
We scale both the width and height of the mesh. The time taken by the ILP solver to solve a rotor matching is measured. The sources of the configured routes are situated at the top left of the device, and the drains are situated at the bottom right of the device. Fig.~\ref{fig:pip3}(b) shows a sub-exponential relation between configuration latency and the width of the PIP hardware.

\section{Conclusion}
We present {\name}, an optical circuit switch using programmable integrated photonics. {\name} uses an ILP model to correctly and optimally solve the switching problem, and finds a set of port matchings to achieve all-to-all connectivity. {\name} demonstrates adequate performance, supporting up to 32 ports with $9\times9$ PIP mesh size.

\

\bibliographystyle{plain}

\end{document}